\documentclass[aps,prl,twocolumn,10pt,showpacs,preprintnumbers,superscriptaddress]{revtex4-2}

\usepackage[dvipsnames,table,xcdraw]{xcolor}

\usepackage{hyperref}
\usepackage{graphicx}  
\usepackage{bm}        
\usepackage{amsmath}   

\begin{document}

\title{Finite temperature magnetic interactions from first principles}

\author{Ravi Kaushik}
    \affiliation{Quantum Materials Theory, Italian Institute of Technology, Via Morego 30, 16163 Genova, Italy}
    \affiliation{Department of Physics, University of Genova, Via Dodecaneso, 33, 16146 Genova GE}
    \author{Ryota Ono}
    \affiliation{National Institute for Materials Science, MANA, 1-1 Namiki, Tsukuba, Ibaraki 305-0044, Japan}
\author{Sergey Artyukhin}
    \email[Correspondence email address: ]{Sergey.Artyukhin@iit.it}
    \affiliation{Quantum Materials Theory, Italian Institute of Technology, Via Morego 30, 16163 Genova, Italy}
\date{\today} 
\graphicspath{{Figures}}
\begin{abstract}  

Density functional theory has demonstrated remarkable predictive power in
calculating magnetic properties at zero temperature. At finite temperatures,
thermally excited phonons may affect magnetism. Efficient ab-initio methods 
to calculate the temperature dependence of magnetic exchange interactions 
are still lacking despite the importance of room temperature magnetism for 
applications. Exchange is controlled by an interplay between metal-ligand hybridization, 
Hubbard repulsion, band gap, interatomic distances and bond angles, all of which change 
with temperature. Here we present a method to calculate the exchange interactions 
at finite temperatures from first principles using only two supercell calculations
and quantify these mechanisms. Changes in bond angles and the band gap are 
identified as a primary factors. In NiO with 180-degree bonds exchange decreases 
with temperature, while in Cr$_2$O$_3$ with the bond angles away from 180 degrees 
the exchange increases by 10\% at room temperature.
\end{abstract}

\maketitle 

Density functional theory and its extensions have been successful in describing magnetic properties of a wide range of material systems at zero temperature. At the same time, many materials of technological importance, such as halide perovskites, molecular magnets, and systems of reduced dimensionality (single and multilayer systems and twisted heterostructures) have strong thermal fluctuations at room temperature. Understanding how temperature affects magnetic interactions is therefore an intriguing challenge. 

\begin{figure}[b]
    \includegraphics[width=\linewidth]{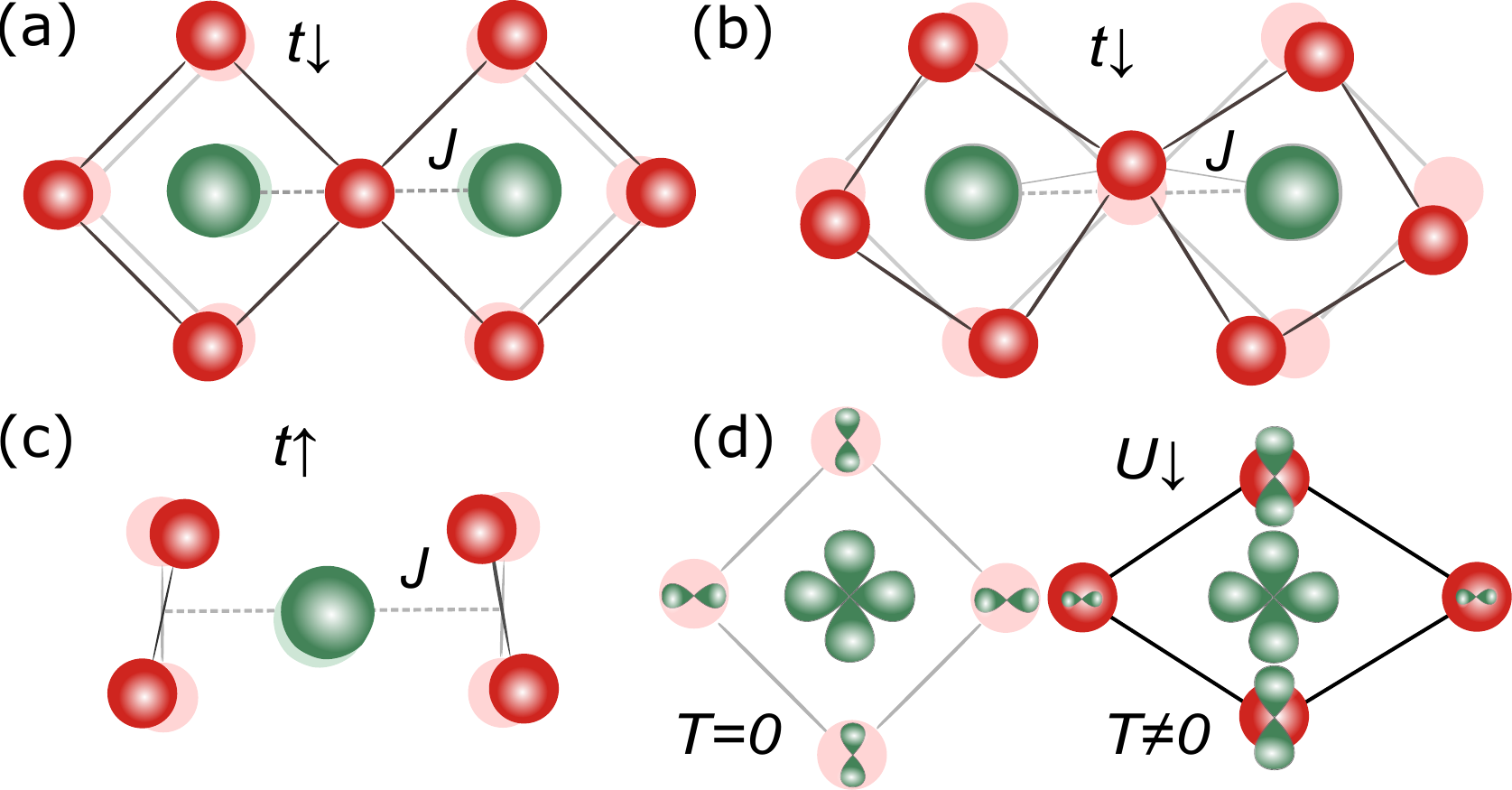}
    \caption{Principal mechanisms driving temperature dependence of magnetic exchange constants: (a) thermal linear expansion elongates the bonds, resulting in the reduction of M-O-M hopping integrals $t_{MO}=t_0-\beta T$ and the exchange constants $J\propto(t_{MO}^2/\Delta)^2/U$; (b) reduction of octahedral tilts with temperature results in enhanced $t_{MO}$ and stronger antiferromagnetic $J$; (c) thermal liberations of magnetic O$_2$ molecules in CsO$_2$ enhance $t$ and $J$ with increasing temperature. (d) Thermal phonons lead to stronger hybridization of Ni with ligands and expansion of the Wannier function (in green) drives the decrease of Hubbard $U$ and increase of $J$ with temperature.}
	\label{fig:mechanisms}
\end{figure}

The majority of materials exhibit a decrease in magnetic interactions as temperature rises \cite{Mankovsky20}. One scenario of such decrease is based on the linear thermal lattice expansion driving the ions apart, and thus reducing the orbital overlap, Fig.~\ref{fig:mechanisms}(a). Since exchange constants $J$ are generally proportional to the overlap integral $t$ squared, $J\propto t^2/U$, they should decrease with increasing temperature \cite{Bramwell1990}. This is observed e.g. in Cu$_2$OSeO$_3$ \cite{Baral23}. However, other mechanisms, driven by phonons, could be at play, and this trend can even be reversed.

One scenario where the exchange may increase with temperature involves a structural transformation from a state of low symmetry to one of higher symmetry with increasing temperature. For example, in transition metal-rare earth perovskites ABX$_3$ the tolerance factor of 1, indicates a perfect match between the sizes of the octahedra BX$_6$ and the A-site ion and results in formation of an ideal cubic structure. A mismatch results in octahedral tilts at low temperatures, illustrated in Fig.~\ref{fig:mechanisms}(b) \cite{Glazer72}. Such tilts reduce the metal-oxygen-metal bond angle away from 180$^\circ$, thus diminishing the exchange constant. At elevated temperatures the tilts disappear, and the bond angle returns to 180$^\circ$, which must boost the magnetic exchange. This could lead to fascinating phenomena, such as re-entrant magnetic ordering, where with increasing $T$ thermal fluctuations destroy the magnetic ordering, which reappears at a higher $T$ due to increased exchange. On the other hand, if no octahedral tilts are present at $T=0$~K, thermal displacements of the ions bend the bonds, thus lowering the exchange constants.

Increase of magnetic exchange due to thermally excited phonons has been found in CsO$_2$ \cite{Felser15}. As illustrated in Fig.~\ref{fig:mechanisms}(c), magnetism here is not derived from the conventional $d$ or $f$ electrons, but rather arises from the $p$ orbitals of oxygen ions forming $\pi^*$ molecular orbitals of O$_2$ dimers.
The increase in exchange constants is thought to be driven by thermal librations of O$_2$ molecules, which cause enhanced orbital overlaps between the O-$p_{x,y}$ orbitals and the Cs-$p_z$ orbitals ~\cite{Magnetogyration}. 

Averaging over the structures, distorted by the presence of thermal phonons, is a major computational bottleneck. In a recent advance Mankovsky et al. calculated exchange constants at non-zero temperature, showing that 14 distorted structures are sufficient to converge the thermal average \cite{Mankovsky20}. 

Motivated by recent modelling of optical absorption spectra using only one structure \cite{Zacharias20}, here we explore the possibility of computing exchange constants at finite temperatures using a small number of first principles supercell calculations, in the spirit of special displacement method by Zacharias and Giustino (ZG)\cite{Zacharias16, Zacharias20}. Our goal is to devise a method to accurately calculate magnetic interactions based on Density Functional Theory (DFT). In doing so, we aim to shed light on the remarkable interplay between temperature, associated ionic displacements and magnetism.

Thermal phonons promote the changes in the electronic structure via electron-phonon coupling. This alters magnetism. Aiming to describe magnetic insulators with magnon frequencies lower than optical phonon ones, we neglect nonadiabatic effects and coherence between magnons and phonons. The case of resonance will be treated in the forthcoming paper. We aim to describe spins with the phonon-averaged Hamiltonian,
\begin{align}
    \mathcal{H} = \frac{1}{2} \sum_{ij} J_{ij} (T) \mathbf{S}_i \cdot \mathbf{S}_j,
\end{align}
where $J_{ij} (T)$ are the temperature-dependent exchange constants between spins $\bm{S}_i$ and $\bm{S}_j$, residing at sites $i$ and $j$, respectively. In a frozen phonon approach, the interactions between electrons and phonons can be approximated by calculating the electronic structure in a supercell with frozen phonons (commensurate with the supercell) that mimic thermal atomic displacements. In the spirit of mean field theory, the most probable phonon amplitudes are chosen so as to minimize the free energy at a given temperature \cite{BornHuang96}, as in ZG method. Thus, we aim to find a frozen-phonon structure, giving the value of $J_{ij}(T)$ which approximates a vibrational average $\langle {J_{ij}}(x_\nu)\rangle_T$. The approach obviates the need for explicit calculations of electron-phonon matrix elements. 

\begin{figure*}[t]
\includegraphics[width=\linewidth]{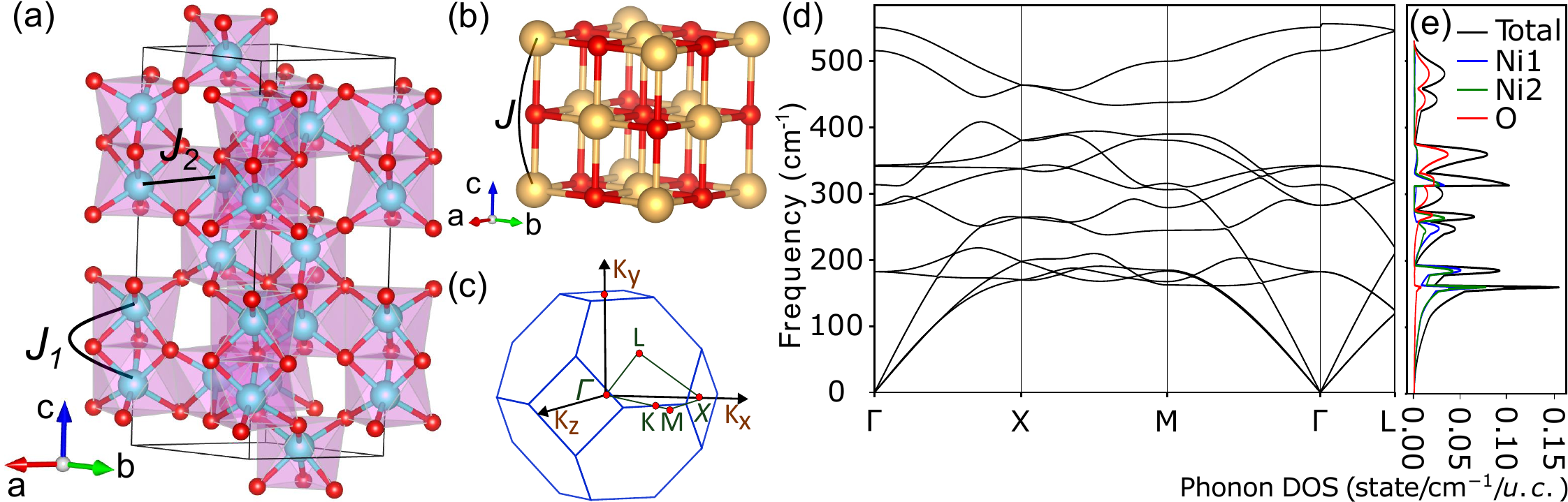}
	\caption{\label{fig:phonons}(a) Crystal structure of Cr$_2$O$_3$  with Cr and oxygen atoms in blue and red, respectively. $J_1$ and $J_2$ are the two nearest-neighbor Heisenberg exchange constants. (b) Crystal structure of NiO. Ni ions are in yellow, while oxygens are in red. Heisenberg exchange $J$ is on a 180$^\circ$ Ni-O-Ni bond. (c) Brillouin zone and a path along which the phonon dispersion is plotted. (d) Calculated phonon dispersion in NiO along the reciprocal space path shown in panel (c). (e) Phonon density of states in NiO.}
\end{figure*}
The requirement of minimizing the free energy, $m_\nu\omega_\nu^2 \sigma_\nu^2=k_{\mathrm{B}}T$ sets the phonon amplitudes $\sigma_\nu$ (with $m_\nu$ and $\omega_\nu$ being the mode effective mass and frequency) but not their relative phases \cite{BornHuang96}. Different phase choices $S_k$ are referred to as thermal lines \cite{monserrat}:
\begin{multline}
\mathcal{T}_\mathbf{S}(T) = (S_1\sigma_{1(\nu,T)}(T), S_2\sigma_{2(\nu,T)}(T), \ldots,\\ S_{3(N-1)}\sigma_{3(N-1)(\nu,T)}(T))
\end{multline}
Without spin-orbit coupling $S_{k}=\pm 1$. $\sigma_{\nu,T}$ refers to r.m.s. phonon amplitudes, satisfying ${m_\nu\omega_\nu^2 \sigma_\nu^2=\hbar\omega_\nu\left(n_{BE}(\hbar\omega_\nu/T)+\frac12\right)}$, resulting in ${\sigma_{\nu,T} = [(2n_{\nu,T} + 1) \ell^2_{\nu}]^\frac{1}{2}}$, where $n_{BE}$ is the Bose-Einstein occupation of the $\nu$-th mode and ${\ell_{\nu} = \left(\hbar/(2M_p\omega_{\nu})\right)^{1/2}}$ is the corresponding zero-point vibrational amplitude, with $M_p$ denoting the proton mass. The thermally averaged $J$ could be computed by averaging over thermal lines. The same can be achieved by considering a supercell calculation which is equivalent to freezing multiple thermal lines in a single primitive cell calculation. Hence larger supercell corresponds to better averaging over thermal lines.

One can choose a particular set of phonon phases, so that all phonons shift the first ion in the $+x$-direction. Adding the phonon displacements with such a phase choice makes the first ion in the frozen phonon structure ``fly away''. These configurations are inaccessible in reality as they have high energy due to anharmonic terms, although such terms are difficult to include. Instead, in the ZG method, a choice of phases $S_k$ is made to minimize the ionic displacements, and thus the unharmonic energy. We do that by performing Monte Carlo simulation over thermal lines $\mathcal{T}$. We start from a random set of $S_k$ that includes equal number of negative and positive signs. Then we iteratively reverse the sign of a randomly chosen phonon and accept the sign flip if it results in a decrease of the target function being the sum of squared ionic displacements in the supercell. We also impose the condition that after the sign update the difference of negative and positive signs must not be greater than 2. We stop the Monte Carlo simulations when the $\{S_k\}$ is no longer changing within 500 steps. That way a representative configuration with appropriate energy is obtained. 
In computing exchange constants, the method only relies on a single DFT calculation for this configuration, therefore enhancing our ability to investigate temperature effects efficiently. The exchange constants are computed using the Katsnelson-Lichtenstein formula \cite{LIECHTENSTEIN198765,Lichtenstein} in the basis of Wannier functions \cite{MOSTOFI20142309,Pizzi_2020}.

The Heisenberg exchange constant for ZG supercell has a form:
\begin{multline}\label{eq:Ja}
J(T)= \underline{J_0}+\sum_\nu \frac{\partial J}{\partial x_\nu} \sigma_{\nu, T}+
\underline{\frac{1}{2} \sum_\nu \frac{\partial^2 J}{\partial x_\nu^2} \sigma_{\nu, T}^2}\\+\frac{3}{3 !} \sum_{\mu \neq \nu} \frac{\partial^3 J}{\partial x_\mu^2 \partial x_\nu} \sigma_{\mu, T}^2 \sigma_{\nu, T}+\frac{1}{3!} \sum_\nu \frac{\partial^3 J}{\partial x_\nu^3} \sigma_{\nu, T}^3\\
+\underline{\frac{6}{4 !} \sum_{\mu \neq \nu} \frac{\partial^4 J}{\partial x_\mu^2 \partial x_\nu^2} \sigma_{\mu, T}^2 \sigma_{\nu, T}^2+\frac{1}{4!} \sum_\nu \frac{\partial^4 J}{\partial x_\nu^4} \sigma_{\nu, T}^4}.
\end{multline}

Therefore, odd-order terms in the phonon amplitudes remain in the expression for $J$. In reality, however, since the phonon amplitudes oscillate on the timescale of slow magnetic dynamics, the first-order variation of $J$ is averaged out. Hence, a single configuration is not suitable for estimating magnetic interactions. It can be proven that the contributions from the modes $k\neq 0$ cancel when averaged over a large supercell, but that does not hold for $k=0$ modes. However, we note that odd-order terms cancel after averaging with the configuration with reversed phonon amplitudes (``$-ZG$").
%
Hence, to compute $J(T)$, we take average over the configurations $ZG$ and $-ZG$:
\begin{equation}
J(T) = \frac{1}{2} \left( \langle 
J[\mathcal{T}_{S_i}] \rangle + \langle J[\mathcal{T}_{-S_i}]\rangle \right).
\end{equation}

\begin{figure*}[t]
	\centering
	\includegraphics[width=\linewidth]{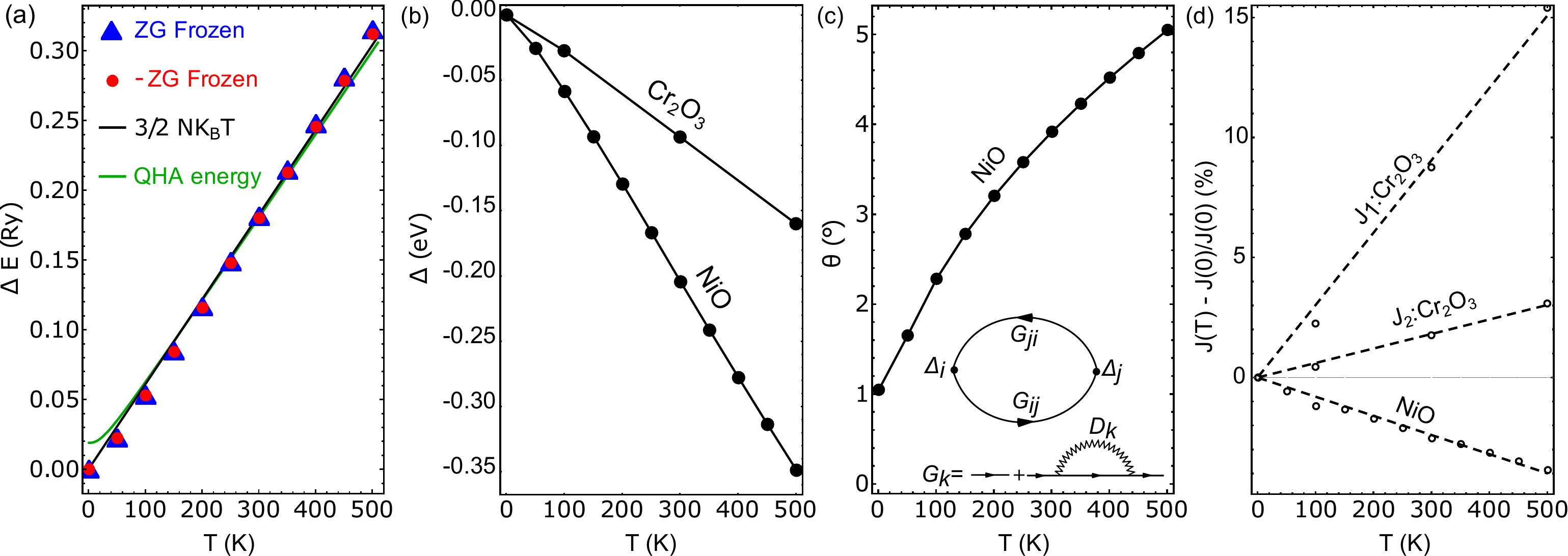}
	\caption{\label{fig:fig4}(a) Increase of total energy with increasing temperature. Red dots and blue triangles show the total energy of +ZG and -ZG configurations. The solid black line is the function $\frac{3}{2}Nk_{\mathrm{B}}T$. (b) Temperature dependent change of the band gap $\Delta$ with temperature. (c) Temperature dependence of the deviation of the average Ni-O-Ni bond angle in NiO from 180$^\circ$. The inset shows the Feynman diagram corresponding to the Katsnelson-Lichtenstein exchange constant $J$ with thermal phonon correction. (d) Temperature dependent change of magnetic exchange for NiO  and Cr$_2$O$_3$. At $T=0$, $J_1=12$, ~meV, $J_2=10$~meV, $J=20$~meV.
    }
\end{figure*}

The odd terms in phonon amplitudes drop and the corrections to the magnetic exchange constant $J(T)$ only contain terms, even in phonon amplitudes (underlined in Eq.~\ref{eq:Ja}), consistent with the Feynman diagram shown in the inset of Fig.~\ref{fig:fig4}(c). This is the essence of the proposed approach.

We benchmark the method on NiO, a paradigmatic antiferromagnetic insulator, and Cr$_2$O$_3$, a well-known magnetoelectric antiferromagnet with a corundum structure, Fig.~\ref{fig:phonons}(a). NiO is known as a charge transfer insulator with a rock salt (NaCl) crystal structure and 180$^\circ$ Ni-O-Ni bonds, as seen in Fig.~\ref{fig:phonons}(b). 

All calculations were performed within the generalized gradient approximation to density functional theory \cite{Ceperley,Perdew} as implemented in the Quantum-Espresso code \cite{QE-2009,QE-2017,QE20}. We used the optimal norm-conserving Vanderbilt (ONCV) pseudopotentials \cite{Hamann13}, generated by Schlipf and Gygi (SG15) \cite{Schlipf15} and PBE exchange-correlation functional\cite{PBE96}. The Kohn-Sham wave functions were expanded in plane-wave basis sets with the kinetic energy cutoff of 80~Ry.  We used the simplified LSDA+U method \cite{dudarev} for the calculation of electronic and magnetic properties to take into account strong Hubbard repulsion in NiO. The rhombohedral unit cell consists of 2 Ni and 2 oxygen atoms. We relaxed the structure with the energy cut-off of 10$^{-8}$~Ry until the forces were reduced below $10^{-4}$~Ry/Bohr. The on-site Coulomb repulsion in the basis of atomic $d$ orbitals is calculated to be $U = 6.42$~eV using the relaxed cell and a dense $q$-grid of $4\times4\times4$ in the HP.x code \cite{timrov2022}. For Cr$_2$O$_3$ we obtained $U=5.25$~eV. In DFT+U calculations we have used the calculated $U$ and the intra-atomic Hund’s rule coupling was taken as $J_0=0.9$~eV \cite{Cococcioni11}. For NiO this gives a band gap of nearly 3.3~eV and a local magnetic moment of 1.72~$\mu_\mathrm{B}$. Phonon calculations used density functional perturbation theory for a rhombohedral cell with 2 formula units (f.u.) accommodating AFM ordering and a $q$-grid commensurate with a cubic cell containing 32 f.u. The interatomic force constants in the Born-von Karman supercell were obtained by Fourier transforming the dynamical matrices \cite{RevModPhys.73.515}. The phonon dispersion and density of states for NiO are shown in Fig.~\ref{fig:phonons}(d,e). The highest optical phonons are at 70 meV.
We selected the best thermal line (set of signs of phonon amplitudes) at 400 K by Monte Carlo minimization of r.m.s. ionic displacements over signs $S_k$. In all our trial runs we iterated until the displacement no longer decreased in 500 updates. We then followed this thermal line, by freezing the phonons into a cubic supercell with 32 formula units and calculating exchange constants for temperatures between 1~K and 501~K in steps of 50~K. Fig.\ref{fig:fig4}(a) shows the linear increase of total energy for frozen phonon structures with $+ZG$ configuration indicated with blue triangles and $-ZG$ with red points. We also compared the increment in energy w.r.t. expected increase from quantum harmonic approximation (QHA), $ \sum_{n,\omega} (n+\frac{1}{2})\hbar\omega$, where $\omega$ are the calculated phonon frequencies (see SI for details). The $E=\frac{3}{2}NK_BT$ fit indicated with the black line in Fig.\ref{fig:fig4}(a) coincides with the QHA curve, except at the lowest temperatures, $T<50$~K.

The main exchange mechanism in NiO and Cr$_2$O$_3$ is superexchange via ligand $p$ orbitals, so the Cr-$d$, Ni-$d$ and O-$p$ orbitals were used for Wannier projections~\cite{Korotin15}. 
$J(T)$ was calculated for each individual temperature with averaging over symmetry-equivalent bonds using Katsnelson-Lichtenstein formula as implemented in the in-house code~\cite{LInvariant}. The 6$\times$6$\times$6 $k$-grid was used for the Fourier transform of the Green's functions to the site representation, and the integration over $\omega$ was performed on a contour, discretized into 4000 points.

In NiO, both for the $+ZG$ and $-ZG$ configurations the magnetic moment decreases slightly, linearly with $T$, and at 500~K the decrease is 0.25\%.
As seen in Fig.~\ref{fig:fig4}(b), the band gap decreases with temperature in both materials, $\Delta^{-1}\partial\Delta/\partial T=-3\cdot 10^{-4}$~K$^{-1}$ and $-7\cdot 10^{-4}$~K$^{-1}$ in Cr2O3 and NiO, respectively. This behavior is typical for semiconductors \cite{Zacharias16} and would alone be expected to drive an increase of $J$. Furthermore, it could lead to an increased screening, which results in a lower Coulomb repulsion on correlated orbitals, i.e. the decrease of Hubbard $U$ (this is schematically indicated by the expansion of the Wannier function in Fig.~\ref{fig:mechanisms}(d)). We have computed $U$ using cRPA approach \cite{timrov2022} and found a slight increase with temperature, $U^{-1}\partial U/\partial T=3\times10^{-5}$~K$^{-1}$ and $2.7\times10^{-4}$ ~K$^{-1}$ for Cr$_2$O$_3$ and NiO respectively. 
The different dependencies of Hubbard $U$ and the gap are likely related to the charge-transfer character of NiO, with the gap separating O-$p$ states and $d$ states of the transition metal ions. 
If $U$ was strongly dependent on $T$, it would be necessary to achieve self-consistency, recomputing the phonon frequencies with the new $U$ value, and harder/softer resulting phonons would drive further change of $U$. Both variations of $U$ and anharmonic phonon interactions modify phonon frequencies. To avoid repeated phonon calculations in a large supercell, a method was proposed to estimate the dynamical matrix of the forces on phonon modes in a distorted structure~\cite{Katsnelson08}. While in rigid NiO and Cr$_2$O$_3$ phonons are rather hard and we do not see a large variation of $J$ in a full self-consistency, in softer materials like NiI$_2$ and CrI$_3$ that could lead to major effects.

Neglecting the weak temperature dependence of $U$ in both materials, we compute $J(T)$ shown in Fig.~\ref{fig:fig4}(d). 
In NiO the bonds at $T=0$ are straight, which maximizes $J$. With increasing temperature, the Ni-O-Ni bond angle deviates from 180$^\circ$, as illustrated in Fig.~\ref{fig:mechanisms}(b) and Fig.~\ref{fig:fig4}(c), thus resulting in a reduced orbital overlap and a decrease in the exchange constant $J$. 
Remarkably, in Cr$_2$O$_3$ we find an opposite behavior of $J(T)$. Here, at low $T$ the M-O-M bonds are not straight, and thermal ionic displacements bring some of them closer to 180$^\circ$, thus driving the increase of average $J$.

We now estimate the magnetic transition  temperature. In the mean field approximation, $T_c=\frac{1}{3k_{\mathrm{B}}S}z (S+1)J$, where z is the lattice coordination number. If $J$ significantly depends on $T$, one must determine $T_c$ self-consistently from the exchange at $T_c$. That gives the equation $J(T)=\frac{3k_{\mathrm{B}}ST}{z(S+1)}$, satisfied for $T_c= 440$~K. If we neglect the temperature dependence in $J(T)$, we obtain $T_c=456$~K instead. Both values agree reasonably well with the experimentally measured $T_N=523$~K\cite{tomlinson55,Hutchings72}. Approximations beyond the mean field may further increase the precision of the $T_c$ estimate \cite{Fischer09}.

We have studied how thermally excited phonons modify spin interactions in solids. Thermal expansion and bond bending can lead to an increase or decrease in the magnetic exchange constants with temperature. A computationally affordable approach based on the special displacement method and the Katsnelson-Lichtenstein formula is devised. The approach is easy to implement on top of any electronic structure code. In NiO, the bending of the Ni-O-Ni bond with temperature results in a decrease of $J$. This is consistent with a general understanding that magnetism weakens with increasing temperature. However, Cr$_2$O$_3$, where bond angles are far from 180$^\circ$, shows a significant increase of $J$ (around 10\% at room temperature). The speed-up given by the method opens the path to high-throughput computation of magnetic exchange constants at finite $T$. The results motivate investigations of temperature effects in soft halide-based magnets, organic-inorganic hybrids, and layered magnetic materials, where thermal effects may be important.

%

\end{document}